\documentclass[10pt]{article}
\usepackage{fullpage}
\usepackage{cite}
\usepackage{graphicx}
\usepackage{amsmath}
\usepackage{amssymb}
\allowdisplaybreaks
\title{}
\date{}

\begin{document}
\bibliographystyle{utphys}
\newcommand{\msbar}{\ensuremath{\overline{\text{MS}}}}
\newcommand{\DIS}{\ensuremath{\text{DIS}}}
\newcommand{\abar}{\ensuremath{\bar{\alpha}_S}}
\newcommand{\bb}{\ensuremath{\bar{\beta}_0}}
\newcommand{\rc}{\ensuremath{r_{\text{cut}}}}
\newcommand{\Nd}{\ensuremath{N_{\text{d.o.f.}}}}
\setlength{\parindent}{0pt}

\newcommand \slsh [1] {\not\!{#1}}

\titlepage
%\begin{flushright}
%Preprint numbers here \\
%\end{flushright}

\vspace*{0.5cm}

\begin{center}
{\Large \bf Diagrammatic insights into next-to-soft corrections}

\vspace*{1cm}
\textsc{C. D. White\footnote{Christopher.White@glasgow.ac.uk}} \\

\vspace*{0.5cm} SUPA, School of Physics and Astronomy, University of Glasgow,\\ Glasgow G12 8QQ, Scotland, UK\\

\end{center}

\vspace*{0.5cm}

\begin{abstract}
We confirm recently proposed theorems for the structure of
next-to-soft corrections in gauge and gravity theories using
diagrammatic techniques, first developed for use in QCD
phenomenology. Our aim is to provide a useful alternative insight into
the next-to-soft theorems, including tools that may be useful for
further study. We also shed light on a recently observed double copy
relation between next-to-soft corrections in the gauge and gravity
cases.
\end{abstract}

\vspace*{0.5cm}

\section{Introduction}
It is well-known that scattering amplitudes in gauge and gravity
theories contain infrared divergences. These arise from the emission
of {\it soft} gluons or gravitons, whose 4-momentum tends to zero. The
remaining {\it hard} particles in the amplitude are then said to obey
the {\it eikonal approximation}, and it can be shown that amplitudes
factorise in this limit. At tree level, for example, the amplitude for
the emission of $n$ hard gluons (momenta $\{p_i\}$) and one soft gluon
(momentum $k$) can be written as
\begin{equation}
{\cal A}_{n+1}(\{p_i\},k)={\cal S}_n^{(0)}{\cal A}_n(\{p_i\});
\quad\quad {\cal S}_n^{(0)}=\sum_{i=1}^n \frac{\epsilon_\mu(k)p_i^\mu}{p_i\cdot k},
\label{Anp1}
\end{equation}
where we neglect the coupling constant and colour factors of the soft
emission for brevity. Here ${\cal A}_n$ is the amplitude for the $n$
hard particles with no additional emission, and $\epsilon_\mu(k)$ the
polarisation vector of the soft gluon. The gravity equivalent of this
is known as Weinberg's soft theorem~\cite{Weinberg:1965nx}, and takes
the form
\begin{equation}
{\cal M}_{n+1}(\{p_i\},k)={\cal S}_{n,{\rm grav.}}^{(0)}{\cal M}_n(\{p_i\});\quad \quad
{\cal S}_{n,{\rm grav.}}^{(0)}=\sum_{i=1}^n \frac{\epsilon_{\mu\nu}(k)p_i^\mu\,p_i^\nu}{p_i\cdot k}.
\label{Mnp1}
\end{equation}

Until recently, much less has been known about the corrections to
eqs.~(\ref{Anp1}, \ref{Mnp1}), upon performing a systematic expansion
in the momentum of the soft gauge boson. Such corrections are known as
next-to-soft, and the hard emitting particles then obey the {\it
  next-to-eikonal} approximation. The phenomenological impact of such
corrections have been studied in
QCD~\cite{Kramer:1996iq,Grunberg:2009yi,Dokshitzer:2005bf,Laenen:2008ux,Almasy:2010wn}, and a systematic attempt to classify them has been made
in~\cite{Laenen:2008gt,Laenen:2010uz,White:2011yy}. The
gravitational consequences of next-to-soft radiation have been
explored in~\cite{Akhoury:2013yua}. \\

An orthogonal recent body of work has explored such contributions from
a more formal point of view. Based on the observation that Weinberg's
soft theorem can be interpreted as a Ward identity associated with BMS
transformations at past and future null
infinity~\cite{Strominger:2013jfa,He:2014laa},
ref.~\cite{Cachazo:2014fwa} conjectured a tree-level next-to-soft
generalisation of eq.~(\ref{Mnp1}), where the subleading soft factor
is given by
\begin{equation}
{\cal S}_{n,{\rm grav.}}^{(1)}=\sum_{i=1}^n \frac{\epsilon_{\mu\nu}(k)p_i^\mu\,k_\rho\,J^{(i)\rho\nu}}{p_i\cdot k}.
\label{NEgrav}
\end{equation}
Here $J^{\rho\nu}$ is the total angular momentum associated with the
hard external leg $i$, and ref.~\cite{Casali:2014xpa} gave an analagous result for gauge theory:
\begin{equation}
{\cal S}_{n}^{(1)}=\sum_{i=1}^n \frac{\epsilon_{\mu}(k)k_\rho\,J^{(i)\mu\rho}}{p_i\cdot k}.
\label{NEgauge}
\end{equation}
These results were subsequently understood from the point of view of
the {\it scattering equations}
of~\cite{Cachazo:2013hca,Cachazo:2013iea} in
ref.~\cite{Schwab:2014xua}, using further symmetry arguments
in~\cite{Larkoski:2014hta,Kapec:2014opa}, and string theoretic ideas in ref.~\cite{Geyer:2014lca,Schwab:2014fia,Bianchi:2014gla}. Higher dimensions were considered
in ref.~\cite{Afkhami-Jeddi:2014fia}, and a holographic description of
the 4-dimensional gravitational theory pursued
in~\cite{Adamo:2014yya}. Possible loop-level corrections to
eqs.~(\ref{NEgrav}, \ref{NEgauge}) have been examined in
refs.~\cite{Bern:2014oka,He:2014bga,Cachazo:2014dia}.\\

The aim of this paper is to explore the above results using Feynman
diagrammatic methods previously developed in
refs.~\cite{Laenen:2008gt,Laenen:2010uz,White:2011yy} (which are
themselves related to the earlier results of
refs.~\cite{Low:1958sn,Burnett:1967km}). Our main motivation is to
clarify how those results are consistent with the recently proposed
theorems. We stress that this analysis is new: whilst
refs.~\cite{Laenen:2008gt,Laenen:2010uz,White:2011yy} and the much
earlier work of refs.~\cite{Low:1958sn,Burnett:1967km} derive partial
results regarding next-to-soft corrections, they do not fully
reproduce the results of eqs.~(\ref{NEgrav}, \ref{NEgauge}).
Secondly, it is nearly always useful to have multiple, equivalent ways
of thinking about a given piece of physics, and we believe that our
point of view may be useful in further studies of the next-to-soft
theorems. Finally, connecting the recent results of
refs.~\cite{Strominger:2013jfa,He:2014laa,Cachazo:2014fwa,Casali:2014xpa,Schwab:2014xua,Larkoski:2014hta,Afkhami-Jeddi:2014fia,Adamo:2014yya,Bern:2014oka,He:2014bga,Cachazo:2014dia,Kapec:2014opa,Geyer:2014lca,Schwab:2014fia,Bianchi:2014gla}
with refs.~\cite{Laenen:2008gt,Laenen:2010uz,White:2011yy} may aid the
ongoing effort to use next-to-eikonal effects to increase the accuracy
of collider predictions. \\

The structure of the paper is as follows. In section~\ref{sec:review}
we briefly review the content of
refs.~\cite{Laenen:2008gt,Laenen:2010uz,White:2011yy}, addressing
next-to-eikonal effects using effective Feynman rules. In
section~\ref{sec:theorem}, we show how these results reproduce the
soft theorem of eq.~(\ref{NEgauge}) for the case of scalar and
fermionic emitting particles. The above references did not consider
external gluons, and we perform this analysis in
section~\ref{sec:rules}. We will confirm the tree-level results of
eq.~(\ref{NEgauge}) for external scalars, fermions and
gluons. Finally, in section~\ref{sec:conclude} we discuss our results
and conclude. Some technical details are presented in the appendix.

\section{Review of necessary concepts}
\label{sec:review}
In this section, we review the results of
refs.~\cite{Laenen:2008gt,Laenen:2010uz,White:2011yy} and related
papers, whose aim is to systematically classify next-to-eikonal
contributions to scattering amplitudes in gauge and gravity
theories. The starting point is to factorise the amplitude into a {\it
  hard function}, which is infrared finite, and a soft function, which
collects all soft singularities~\footnote{One must also include {\it
    jet functions} to keep track of collinear singularities. For the
  purposes of the present paper, however, we may implicitly absorb the
  jets into the hard function, as in
  refs.~\cite{Laenen:2008gt,Laenen:2010uz,White:2011yy}.}. Such a
factorisation is well-known (see e.g. ref.~\cite{Gardi:2009zv} for a
review in QCD, and refs.~\cite{Naculich:2011ry,Akhoury:2011kq} for
gravity). However, refs.~\cite{Laenen:2008gt,White:2011yy} generalised
the soft function to include next-to-soft radiative corrections. The
method proceeded by writing the propagators for the external particles
in a background soft gauge field as first-quantised path
integrals~\cite{Strassler:1992zr,vanHolten:1995ds}, which can be
evaluated perturbatively. The leading term in this expansion is the
eikonal approximation, in which external particles do not recoil, and
change only by a (Wilson-line) phase~\cite{Korchemsky:1992xv}. The
first subleading term describes the emission of next-to-soft gauge
bosons, which are completely external to the hard
interaction. Reference~\cite{Laenen:2010uz} rederived the same results
via a systematic expansion of Feynman diagrams to all orders in
perturbation theory, and also checked the resulting formalism by
reproducing known next-to-eikonal logarithms in Drell-Yan
production. These are not the only sources of next-to-soft
correction. As refs.~\cite{Laenen:2008gt,Laenen:2010uz,White:2011yy}
explain in detail, one must also worry about soft gluon emissions
which originate from inside the hard interaction. \\

\begin{figure}[t]
\begin{center}
\scalebox{0.7}{\includegraphics{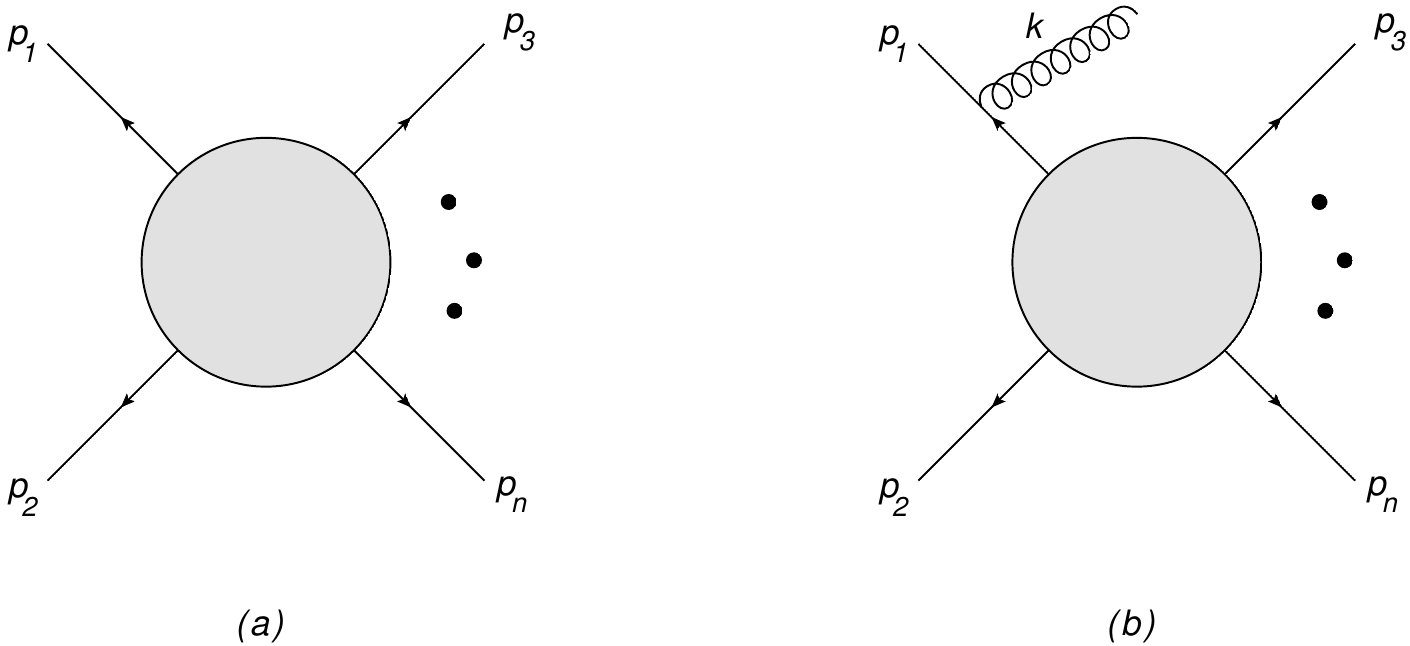}}
\caption{(a) A hard interaction which produces $n$ particles; (b)
  Emission of an eikonal gluon from an external leg.}
\label{fig:amp}
\end{center}
\end{figure}

Let us illustrate how the results apply to the present context, namely
that of dressing an amplitude for the emission of $n$ hard particles
by an additional (next-to-) soft emission. One starts with a hard
interaction such as that shown in figure~\ref{fig:amp}(a). The leading
soft singularities come from dressing all external legs (momenta
$\{p_i\}$) with a soft gauge boson (momentum $k$), whose emission is
described by an eikonal Feynman rule. This is shown in
figure~\ref{fig:amp}(b), and the kinematic parts of the eikonal
Feynman rules for Yang-Mills theory and gravity are
\begin{equation}
\frac{p_i^\mu}{p_i\cdot k}\qquad {\rm and}\qquad 
\frac{p_i^\mu\,p_i^\nu}{p\cdot k}
\label{eikrules}
\end{equation}
respectively. This clearly leads to the soft factors of
eqs.~(\ref{NEgrav}, \ref{NEgauge}), and at leading soft level one need only
worry about the external emission of soft gluons. In Feynman diagram
language, this can be understood by the fact that a soft gluon landing
inside the hard interaction cuts squares an offshell propagator, which
dampens the infrared singular behaviour. In more physical terms, a
soft gluon has an infinite Compton wavelength, and thus cannot resolve
the substructure of the hard interaction. For the same reason, the
above eikonal Feynman rules are independent of the spin of the hard
emitting particles. \\

\begin{figure}
\begin{center}
\scalebox{0.7}{\includegraphics{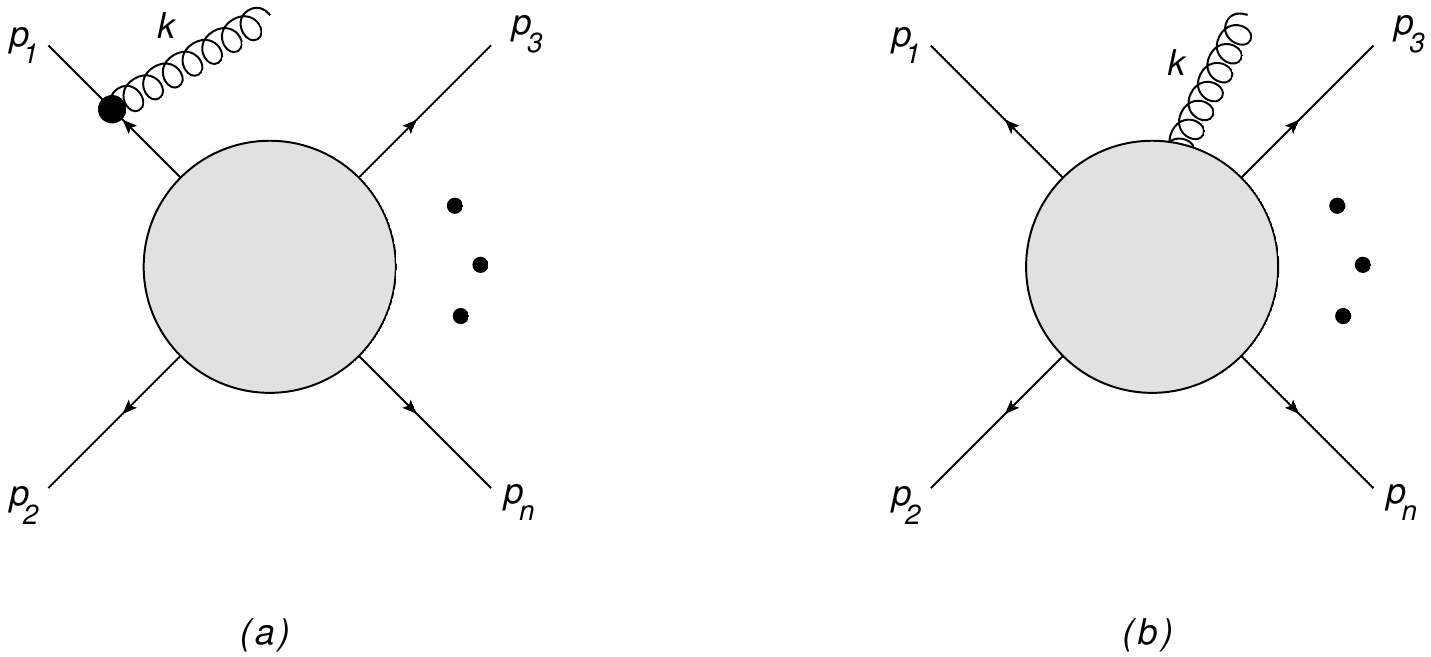}}
\caption{(a) External emission of a next-to-soft gluon; (b)
  Internal emission of a soft gluon.}
\label{fig:NEamp}
\end{center}
\end{figure}

At next-to-soft level, there are two types of contribution. Firstly,
there are next-to-soft gluon emissions external to the hard
interaction, as shown in figure~\ref{fig:NEamp}(a). These emissions
are described by next-to-eikonal (NE) Feynman rules which, unlike the
purely soft limit, depend on the spin of the emitting
particles. External fermions and scalars were considered in Yang-Mills
theory in refs.~\cite{Laenen:2008gt,Laenen:2010uz}; scalars only were
considered in the gravity study of ref.~\cite{White:2011yy}.
In Yang-Mills theory, the NE Feynman rules for emission of a
(potentially off-shell) gluon from a scalar and fermion are
\begin{equation}
V^\mu_{\rm scal.}=\frac{k^\mu}{2p_i\cdot k}-\frac{k^2\,p^\mu}{2(p_i\cdot k)^2};\qquad V^\mu_{\rm
  ferm.}=V^\mu_{\rm scal.}-\frac{ik_\nu\,\Sigma^{\mu\nu}}{p_i\cdot
  k},
\label{1gv}
\end{equation}
where 
\begin{equation}
\Sigma^{\mu\nu}=\frac{i}{4}\left[\gamma^\mu,\gamma^\nu\right]
\label{sigmadef}
\end{equation}
is the generator of Lorentz transformations. There are also NE Feynman
rules describing the correlated emission of a pair of soft gluons. The
result for emission from an external scalar, for example, is
\begin{equation}
R^{\mu\nu}=\frac{p\cdot k_1\, k_2^\mu\, p^\nu\,+p\cdot k_2\,k_1^\nu\,p^\mu-p^\mu\,p^\nu k_1\cdot k_2 -\eta^{\mu\nu}p\cdot k_1\,p\cdot k_2}{p\cdot k_1\,p\cdot k_2\,p\cdot (k_1+k_2)},
\label{Rmunudef}
\end{equation}
where $k_1^\nu$ and $k_2^\mu$ are the soft gluon 4-momenta. An
additional contribution arises for external fermions, again involving
the generator of Lorentz
transformations~\cite{Laenen:2008gt,Laenen:2010uz}. Similar Feynman
rules have been obtained for scalar external legs emitting
next-to-soft gravitons in ref.~\cite{White:2011yy}, where again there
are one and two-graviton vertices.\\

The second type of contribution at next-to-eikonal level arises from
the {\it internal emission} of a soft boson from inside the hard
interaction. Such contributions violate the factorisation of an
amplitude into hard and soft parts; physically, this corresponds to
the emitted soft boson being able to resolve the finite size of the
hard interaction. However, gauge invariance fixes these contributions
in terms of derivatives with respect to the external momenta, acting
on the hard interaction with no additional emission. This result was
first derived by Low~\cite{Low:1958sn} for scalar particles, and
generalised to fermions by Burnett and Kroll~\cite{Burnett:1967km}. A
further important generalisation, to massless external particles, was
carried out by Del Duca~\cite{DelDuca:1990gz}. The internal emission
contributions decompose into a part which is independent of the spin
of the external legs, and an additional spin-dependent piece. The
latter is associated only with the hard collinear region, and need not
concern us here. The former can be written, in the present notation,
as
\begin{equation}
{\cal A}_{n+1}^{\rm int.}=\sum_i \left(\frac{p_{i\mu}}{p_i\cdot k}k_{\nu}\frac{\partial}{\partial p_i^\nu}-\frac{\partial}{\partial p_i^\mu}\right){\cal A}_n(\{p_i\}).
\label{internal}
\end{equation}
In the path integral approach of ref.~\cite{Laenen:2008gt}, the first
term arises from the non-zero initial position of each hard external
line, thus making clear that this contribution arises due to the
non-trivial spatial extent of the hard interaction. The analogous
result for gravity reads~\cite{White:2011yy}
\begin{align}
{\cal M}_{n+1}^{\rm int.}=\sum_j \left(\frac{p_{j\mu}p_{j\nu}}{{p_j\cdot k}} k^\sigma\frac{\partial}{\partial p_j^\sigma}-p_{j\mu}\frac{\partial}{\partial p_j^\nu}\right){\cal M}_n(\{p_n\}).
\label{internal2}
\end{align}
To summarise, next-to-soft contributions at a given order can be
calculated by combining external and internal emission
contributions. The former are described using effective NE Feynman
rules, whereas the latter obey the iterative formulae of
eqs.~(\ref{internal}, \ref{internal2}). \\

\section{The next-to-soft theorem for scalars and fermions}
\label{sec:theorem}

In this section, we demonstrate how the tree-level soft theorems of
eq.~(\ref{NEgrav}, \ref{NEgauge}) are reproduced by the results of
refs.~\cite{Laenen:2008gt,Laenen:2010uz,White:2011yy}, beginning with
scalar emitting particles in Yang-Mills theory. As described above,
one must combine external and internal emission contributions. The
former can be obtained from the NE Feynman rules of eqs.~(\ref{1gv},
\ref{Rmunudef}). The two-gluon vertex does not contribute, owing to
the fact that we are taking only one gluon soft, and remain at tree
level. Furthermore, both terms in the 1-gluon vertex of
eq.~(\ref{1gv}) vanish: the first due to contraction with a physical
polarisation tensor obeying $k^\mu\epsilon_\mu(k)=0$, and the second
due to the onshellness of the emitted gluon ($k^2=0$). Thus, only the
internal emission contributions are necessary, which may be
rewritten as
\begin{align}
{\cal A}_{n+1}^{\rm int.}&=\frac{k^\nu}{p_i\cdot k}\left(p_{i\mu}\frac{\partial}{\partial p_i^\nu}-p_{i\nu}\frac{\partial}{\partial p_i^\mu}\right){\cal A}_n\notag\\
&=-\frac{ik^\nu\,L^{(i)}_{\mu\nu}}{p_j\cdot k}{\cal A}_n,
\label{gauge2}
\end{align}
where we have introduced the orbital angular momentum tensor for the
$i^{\rm th}$ particle:
\begin{equation}
L^{(i)}_{\mu\nu}=x_{i\mu}\, p_{i\nu}-x_{i\nu}\, p_{i\mu}=i\left(p_{i\mu}\frac{\partial}{\partial p_i^\nu}-p_{i\nu}\frac{\partial}{\partial p_i^\mu}\right).
\end{equation}
For a scalar particle, the orbital angular momentum is equal to the
total angular momentum, $L^{(i)}_{\mu\nu}=J^{(i)}_{\mu\nu}$, and thus we
have indeed reproduced the next-to-soft theorem of eq.~(\ref{NEgauge}).  \\

Considering now fermionic emitting particles, the internal emission
contribution will be the same as for the scalar case. However, there
is now a non-zero external emission contribution, due to the magnetic
moment term in eq.~(\ref{1gv}). The total next-to-soft contribution is
then
\begin{align}
{\cal A}_{n+1}&=\frac{k^\nu}{p_i\cdot k}\left(p_{i\mu}\frac{\partial}{\partial p_i^\nu}-p_{i\nu}\frac{\partial}{\partial p_i^\mu}-i\Sigma_{\mu\nu}\right){\cal A}_n\notag\\
&=-\frac{ik^\nu\,(L^{(i)}_{\mu\nu}+\Sigma_{\mu\nu})}{p_j\cdot k}{\cal A}_n,
\label{gauge3}
\end{align}
One may recognise the bracketed factor in the numerator as the sum of
the orbital and spin angular momentum of the $i^{\rm th}$ particle,
and thus one finds
\begin{align}
{\cal A}_{n+1}&=-\frac{ik^\nu\,J^{(i)}_{\mu\nu}}{p_j\cdot k}{\cal A}_n
\label{gauge4}
\end{align}
as before.\\

We may also examine the case of gravity, and the effective Feynman
rules for scalar emitting particles were first derived in
ref.~\cite{White:2011yy}. However, that paper defined the graviton in
terms of the metric $g_{\mu\nu}$ and its determinant $g$ via
\begin{equation}
\sqrt{-g}\,g_{\mu\nu}=\eta_{\mu\nu}+\kappa h_{\mu\nu}
\label{grav1}
\end{equation}
rather than the more conventional (in high energy physics) choice
\begin{equation}
g_{\mu\nu}=\eta_{\mu\nu}+\kappa h_{\mu\nu}.
\label{grav2}
\end{equation}
Next-to-eikonal Feynman rules for the definition of eq.~(\ref{grav2})
are derived here in appendix~\ref{app:gravrules}, where we also derive
the rules for graviton emission from fermions. The resulting
one-graviton vertices for the emission of a graviton of momentum $k$
from a hard line of momentum $p$ are
\begin{align}
V^{\mu\nu}_{\rm scal.}&=-\frac{p^\mu\,p^\nu\,k^2}{2(p\cdot k)^2}+\frac{k^{(\nu}p^{\mu)}}{2p\cdot k}-\frac{\eta^{\mu\nu}}{2};\notag\\
V^{\mu\nu}_{\rm ferm.}&=V^{\mu\nu}_{\rm scal.}-\frac{ik_\rho\Sigma^{\rho(\mu}p^{\nu)}}{2p\cdot k}.
\label{gravrule3}
\end{align}
As in the Yang-Mills case, all scalar-like external emission
contributions vanish. This is due to onshellness of the emitted
graviton, and contraction with a physical, traceless polarisation
tensor for the graviton:
\begin{equation}
\epsilon_{\mu\nu}(k)k^\mu=\epsilon_{\mu\nu}(k)k^\nu=\epsilon_{\mu\nu}(k)\eta^{\mu\nu}=0.
\label{polargrav}
\end{equation}
For the scalar case, then, the only next-to-soft contributions arise
from internal emissions, which are given
by~\cite{White:2011yy}\footnote{Note that ref.~\cite{White:2011yy}
  contains a number of typos, which have been fixed in
  eq.~(\ref{intgrav}).} 
\begin{align}
{\cal M}_{n+1}^{\rm int.}&=
\sum_j \left(\frac{p_{j\mu}p_{j\nu}}{{p_j\cdot k}} k^\rho\frac{\partial}{\partial p_j^\rho}-p_{j\mu}\frac{\partial}{\partial p_j^\nu}\right){\cal M}_n(\{p_i\})\notag\\
&=\sum_j\frac{p_{j\mu}\,k^\rho}{p_j\cdot k}\left(p_{j\nu}\frac{\partial}{\partial p_j^\rho}-p_{j\rho}\frac{\partial}{\partial p_j^\nu}\right){\cal M}_n(\{p_i\})\notag\\
&=\sum_j \frac{-ip_{j\mu} k^\rho L^{(j)}_{\rho\nu}}{p_j\cdot k}{\cal M}_n(\{p_n\}).
\label{intgrav}
\end{align}
Given that the orbital angular momentum is the total angular momentum
in this case, this is the next-to-soft theorem of
eq.~(\ref{NEgrav}). For fermionic emitting particles, one must add the
additional external emission contribution from eq.~(\ref{gravrule3}),
which gives a total next-to-soft amplitude
\begin{align}
{\cal M}_{n+1}&=\sum_j \frac{-ip_{j\mu} k^\rho (L^{(j)}_{\rho\nu}+\Sigma_{\rho\nu})}{p_j\cdot k}{\cal M}_n(\{p_n\})\notag\\
&=\sum_j \frac{-ip_{j\mu} k^\rho J^{(j)}_{\rho\nu}}{p_j\cdot k}{\cal M}_n(\{p_n\}),
\label{intgrav2}
\end{align}
where we have again recognised the total angular momentum as the sum
of the orbital and spin contributions. This is indeed the theorem of
eq.~(\ref{NEgrav}).\\

Before moving on, it is interesting to note that if one neglects the
term in $\eta^{\mu\nu}$ in eq.~(\ref{gravrule3}) (which in any case
gives zero when contracted with a physical polarisation tensor), each
term in the NE one-graviton vertex has the form of a gauge-theory
eikonal Feynman rule multiplying a gauge-theory NE rule. This is
strongly reminiscent of the double copy between gravity and gauge
theory of refs.~\cite{Bern:2010ue,Bern:2010yg}, and indeed it has
already been noted in ref.~\cite{He:2014bga} (see after eq.~(2.15) of
that paper) that the next-to-soft theorems in gauge theory and gravity
theories appear to be related by the double copy. The similarity
between the NE Feynman rules observed here is itself a generalisation
of the fact that the eikonal Feynman rules in gauge and gravity
theories are related by the double copy. This was explored in detail
in ref.~\cite{Oxburgh:2012zr}, where it was shown that matching up the
infrared singularities of Yang-Mills theory and gravity provides
all-order evidence for the double copy conjecture.

\section{The next-to-soft theorem for gluons}
\label{sec:rules}

In the previous section, we saw how to derive the next-to-soft
theorems for external scalars and fermions from a Feynman diagrammatic
treatment. In this section, we consider the Yang-Mills result for the
case of external gluons. These were not considered in
refs.~\cite{Laenen:2008gt,Laenen:2010uz}, and so we must derive the
relevant NE Feynman rules. Our derivation will be analagous to that
carried out in appendix~\ref{app:gravrules}, although we will consider
the sum of internal and emission contributions at the outset.\\

\begin{figure}
\begin{center}
\scalebox{0.8}{\includegraphics{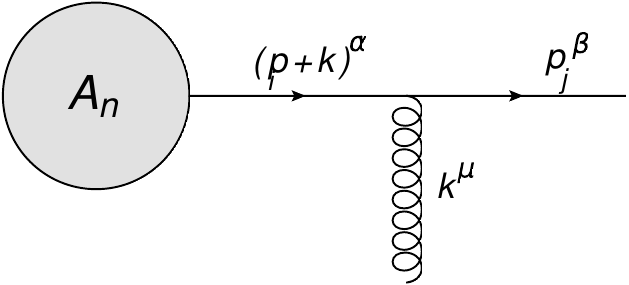}}
\caption{Emission of a single gluon from the $j^{\rm th}$ external
  leg of the hard amplitude ${\cal A}_n$.}
\label{fig:gluonamp}
\end{center}
\end{figure}

Let us start with the $n$-point hard amplitude of
figure~\ref{fig:amp}(a), and consider all places in which one may add
an additional gluon. As described already above, we may emit this
either from an external line, or from inside the hard
interaction. Using the usual Feynman rules of QCD, one then finds that
the $(n+1)$ amplitude is given by
\begin{align}
{\cal A}_{n+1}(\{p_i\},k)&=\left({\cal A}_{n+1}^{\mu,{\rm int.}}-\sum_{j=1}^n {\bf T}_j\,{\cal A}^\alpha_n(p_j+k) \frac{\left[\eta^{\alpha\beta}(-k-2p_j)^\mu+\eta^{\alpha\mu}(2k+p_j)^\beta+\eta^{\beta\mu}(p_j-k)^\alpha\right]\epsilon_\beta(p)}{(p_j+k)^2}\right)\notag\\
&\quad\times\epsilon_\mu(k).
\label{gluonamp1}
\end{align}
Here the first term collects the internal emission contributions, and
${\bf T}_j$ is a colour generator associated with an external emission
on line $j$, which we keep track of for reasons that will become
clear. We have also used the notation $A^\alpha_{n}(p_j+k)$ to denote
the $n$-particle amplitude, but where the hard momentum $p_j$ has been
replaced by $p_j+k$, and where $\alpha$ is the Lorentz index of the
$j^{\rm th}$ external gluon line. The emission from this line is then
as shown in figure~\ref{fig:gluonamp}.
One may expand eq.~(\ref{gluonamp1}) up to first subleading order in
the momentum $k$, and simplify the result using the transverse nature
of the polarisation vector
\begin{equation}
\epsilon_\beta(p_j)p_j^\beta=0
\label{transid}
\end{equation}
as well as the Ward identity
\begin{equation}
{\cal A}_n^\alpha(p_j)p_{j\alpha}=0.
\label{Ward}
\end{equation}
The result is
\begin{align}
{\cal A}_{n+1}(\{p_i\},k)&=\left\{{\cal A}_{n+1}^{\mu,{\rm int.}}+\sum_{j=1}^n{\bf T}_j\,\left[\left[\eta^{\alpha\beta}\left(\frac{p_j^\mu}{p_j\cdot k}+\frac{k^\mu}{2p_j\cdot k}\right)+\frac{k^\alpha\eta^{\mu\beta}}{2p_j\cdot k}-\frac{k^\beta\eta^{\alpha\mu}}{p_j\cdot k}\right]{\cal A}_{n\alpha}(p_j)\right.\right.\notag\\
&\left.\left.\quad+\left(\frac{p_j^\mu}{p_j\cdot k}\eta^{\alpha\beta}k^\sigma-\frac{\eta^{\mu\beta}p_j^\alpha k^\sigma}{2p_j\cdot k}\right)\frac{\partial {\cal A}_{n\alpha}(p_j)}{\partial p_j^\sigma}\right]\right\}.
\label{gluonamp2}
\end{align}
One may simplify this result still further by noting that
eq.~(\ref{Ward}) implies
\begin{equation}
\frac{\partial}{\partial p^\sigma}\left[p^\alpha {\cal A}_{n\alpha(p)}\right]=0
\quad\Rightarrow\quad p^\alpha\frac{\partial {\cal A}_{n\alpha}(p)}{\partial p^\sigma}=-\delta^\alpha_\sigma {\cal A}_{n\alpha}(p).
\label{Ward2}
\end{equation}
One then finds
\begin{align}
{\cal A}_{n+1}(\{p_i\},k)&=\left\{{\cal A}_{n+1}^{\mu,{\rm int.}}+\sum_{j=1}^n{\bf T}_j\,\left[\eta^{\alpha\beta}\left(\frac{p_j^\mu}{p_j\cdot k}+\frac{k^\mu}{2p_j\cdot k}\right){\cal A}_{n\alpha}(p_j)+\eta^{\alpha\beta}\frac{p_j^\mu k^\sigma}{p_j\cdot k}\frac{\partial {\cal A}_{n\alpha}(p_j)}{\partial p_j^\sigma}\right.\right.\notag\\
&\left.\left.\quad-{\cal A}_{n\alpha}\frac{k_\sigma}{p_j\cdot k}\left(\eta^{\sigma\beta}\eta^{\alpha\mu}-\eta^{\sigma\alpha}\eta^{\beta\mu}\right)\right]\epsilon_\beta(p_j)\right\}\epsilon_\mu(k).
\label{gluonamp3}
\end{align}
Let us now interpret this result. Firstly, the first and third terms
in the square bracket can be written in the form
\begin{equation}
\sum_{j=1} {\mathbf T_i} {\cal A}_{n\alpha}\left[V^\mu_{\rm vec.}\right]_{\alpha\beta}\,\epsilon_\beta(p),
\label{Vmuvec}
\end{equation}
where the effective Feynman rule for the external emission of a gluon
up to next-to-soft order is
\begin{equation}
V^\mu_{\rm vec.}=\frac{p_j^\mu}{p_j\cdot k}+\frac{k^\mu}{2p_j\cdot k}-i\frac{k_\sigma\,M^{\sigma\mu}}{p_j\cdot k},
\label{Vmuvec2}
\end{equation}
and we have recognised the generator of the Lorentz group which acts
on vector fields
\begin{equation}
M^{\mu\nu}_{\alpha\beta}=i\left[\delta^\mu_\alpha\delta^\nu_\beta-\delta^\nu_\alpha\delta^\mu_\beta\right].
\label{Mdef}
\end{equation}
The first term in eq.~(\ref{Vmuvec2}) is the eikonal Feynman rule for the emission of a gluon from an external line; the remaining terms produce the next-to-soft external emission contributions. As in the fermion case, there is a spin-independent contribution, and a part involving the spin angular momentum of each external gluon. \\

The remaining term in the square bracket in eq.~(\ref{gluonamp3}) can
be related to the internal emission contribution. One may see this by
writing
\begin{equation}
{\cal A}_{n+1}(\{p_i\},k)={\cal A}_{n+1}^\mu \epsilon_\mu(k)
\label{Amudef}
\end{equation}
on the left-hand side, and applying the Ward identity 
\begin{equation}
k_\mu{\cal A}_{n+1}^\mu=0
\label{Ward3}
\end{equation}
which according to the right-hand side of eq.~(\ref{gluonamp3})
implies
\begin{equation}
k_\mu{\cal A}_{n+1}^{\mu,{\rm int.}}+\sum_{j=1}^n {\mathbf T}_j\, \left[{\cal A}_{n\alpha}(p_j)+k_\mu\frac{\partial{\cal A}_{n\alpha}(p_j)}{\partial p_j^\mu}\right]\epsilon_\alpha(p_j)=0.
\end{equation}
The first term in the square brackets cancels by colour conservation
\begin{equation}
\sum_{j=1}^n {\mathbf T}_j=0,
\label{colcon}
\end{equation}
and one thus obtains
\begin{equation}
{\cal A}_{n+1}^{\mu,{\rm int.}}=-\sum_{i=1}^n {\mathbf T}_j \frac{\partial{\cal A}_{n\alpha}}{\partial p_j^\mu}\epsilon_\alpha(p_j).
\label{Anp1int}
\end{equation}
This is in fact a rederivation of part of the internal emission
contribution that we have already quoted for the scalar case in
eq.~(\ref{gauge2}), and is similar to the original analysis by
Low~\cite{Low:1958sn}. The result is incomplete for massless external
particles, however, as explained in detail by Del Duca in
ref.~\cite{DelDuca:1990gz}. Here, the loophole in the above derivation
is that we have not carefully separated out collinear singularities,
implicitly absorbing jet functions into the hard function~\footnote{It
  is for this reason that the original theorems by Low, Burnett and
  Kroll~\cite{Low:1958sn,Burnett:1967km} do not fully reproduce the
  next-to-soft theorems of eqs.~(\ref{NEgrav}, \ref{NEgauge}), as
  alluded to in the introduction.}.  A more careful analysis leads to
the full result of eq.~(\ref{gauge2}), which is independent of the
spin of the emitting particles (additional spin-dependent
contributions are associated with the hard collinear region, and thus
absent in the soft expansion~\cite{DelDuca:1990gz}).\\

One now obtains the sum of all next-to-soft contributions at
tree-level by combining the external and internal emission
contributions, as for the scalar and fermion cases. The only surviving
NE contribution from eq.~(\ref{Vmuvec2}) is the spin-dependent piece,
and one finds~\footnote{Note that we have now left colour matrices
  implicit, consistent with the notation throughout the rest of the
  paper.}
\begin{align}
{\cal A}_{n+1}&=-\frac{ik^\nu\,(L^{(i)}_{\mu\nu}+M_{\mu\nu})}{p_j\cdot k}{\cal A}_n\notag\\
&=-\frac{ik^\nu\,J^{(i)}_{\mu\nu}}{p_j\cdot k}{\cal A}_n,
\label{gluontot}
\end{align}
thus reproducing the next-to-soft theorem of eq.~(\ref{NEgauge}).\\

In this section, we have seen how similar methods to those used in
refs.~\cite{Laenen:2008gt,Laenen:2010uz} can be used to derive the
next-to-soft theorem of eq.~(\ref{NEgauge}) at tree-level. A similar
analysis could be carried out for gravity. This would be much more
cumbersome, however, due to the lengthy form of the expression for the
three-graviton vertex when written in a helicity-independent
form. 

\section{Conclusion}
\label{sec:conclude}
There has recently been a flurry of
attention~\cite{Strominger:2013jfa,He:2014laa,Cachazo:2014fwa,Casali:2014xpa,Schwab:2014xua,Larkoski:2014hta,Afkhami-Jeddi:2014fia,Adamo:2014yya,Bern:2014oka,He:2014bga,Cachazo:2014dia,Kapec:2014opa,Geyer:2014lca,Schwab:2014fia,Bianchi:2014gla}
focussing on the behaviour of scattering amplitudes in the
next-to-soft approximation, in which a single external particle is
taken to have a small, but non-zero, momentum. The structure of such
contributions is formally interesting in its own right, but also has
distinct practical applications in improving collider physics
predictions~\cite{Kramer:1996iq,Grunberg:2009yi,Dokshitzer:2005bf,Laenen:2008ux,Almasy:2010wn}. In
this paper, we have examined the recently conjectured next-to-soft
theorems of eqs.~(\ref{NEgrav}, \ref{NEgauge}) using diagrammatic
methods developed in
refs.~\cite{Laenen:2008gt,Laenen:2010uz,White:2011yy}. There are a
number of motivations for this. Firstly, our approach provides a
useful alternative view on how the next-to-soft theorems arise,
especially given that it is manifestly independent of the helicities
of the emitting particles, and also the space-time dimension. It is
interesting, for example, to see how the orbital and spin angular
momentum contributions combine to create the coupling to the total
angular momentum of each external leg. We also saw that the NE Feynman
rules for gravity and gauge theory have a (partial) double-copy
structure, which almost certainly underlies the observation made in
ref.~\cite{He:2014bga} that the next-to-soft factors in
eq.~(\ref{NEgrav}, \ref{NEgauge}) have this property (analogous to the
strictly eikonal analysis of ref.~\cite{Oxburgh:2012zr}).\\

We hope that our study clarifies the relationship between the recent
studies on next-to-soft theorems, and previous work in the
literature. Moreover, we believe that the diagrammatic techniques
discussed herein may prove very useful in further examination of
e.g. loop corrections. In particular, at loop level one has to worry
about the two-gluon (or graviton) vertex (eq.~(\ref{Rmunudef}) and its
generalisations), such that one gluon is real and the other
virtual~\cite{NEinprep}. Our techniques may also prove useful for
investigating the phenomenological consequences of next-to-soft
behaviour. Work in this regard is ongoing.\\

{\bf Addendum}\\

In the final stages of this paper, the author became aware of the
recent refs.~\cite{Broedel:2014fsa,Bern:2014vva}, which also address
how to systematically classify next-to-soft corrections.

\section*{Acknowledgments}
CDW is supported by the UK Science and Technology Facilities Council
(STFC). He thanks Duff Neill for correspondence that initiated this
study, and is greatly indebted to Domenico Bonocore, Eric Laenen,
Lorenzo Magnea, Stacey Melville and Leonardo Vernazza for ongoing
collaboration on related topics. In addition, he is grateful to
Christoph Englert, Einan Gardi, Grisha Korchemsky, David Miller and
Donal O'Connell for discussions.

\appendix

\section{NE Feynman rules for graviton emission}
\label{app:gravrules}
\begin{figure}
\begin{center}
\scalebox{0.8}{\includegraphics{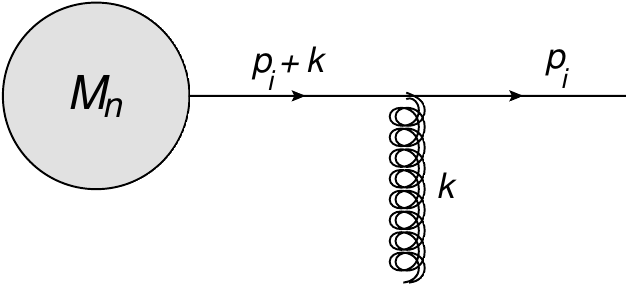}}
\caption{Emission of a single graviton from the $i^{\rm th}$ external
  leg of the hard amplitude ${\cal M}_n$.}
\label{fig:ampleg}
\end{center}
\end{figure}

In this appendix, we calculate the effective Feyman rules up to
next-to-eikonal order for the emission of gravitons from external
scalars or fermions. We begin with the diagram of figure~\ref{fig:ampleg}, which shows a
single leg of the hard amplitude (momentum $p_i$) emitting a graviton
(momentum $k$). Examining first the case of a scalar emitter, one may
combine the scalar propagator and graviton-scalar vertex (see
e.g.~\cite{Veltman:1975vx}) to get
\begin{equation}
{\cal M}_n\left[\frac{p_i^{(\mu}(p_i+k)^{\nu)}-\eta^{\mu\nu}p_i\cdot(p_i+k)}{(p_i+k)^2}\right],
\label{gravrule1}
\end{equation}
where we have used the notation
\begin{displaymath}
a^{(\mu}b^{\nu)}=a^\mu\,b^\nu+a^\nu\,b^\mu
\end{displaymath}
and neglected a factor of the gravitational coupling $\kappa/2$ for
brevity. Expanding the expression~(\ref{gravrule1}) up to first
subleading order in the emitted graviton momentum gives
\begin{equation}
{\cal M}_n\left[\frac{p_i^\mu\,p_i^\nu}{p_i\cdot k}-\frac{p_i^\mu\,p_i^\nu\,k^2}{2(p_i\cdot k)^2}+\frac{k^{(\nu}p^{\mu)}}{2p\cdot k}-\frac{\eta^{\mu\nu}}{2}\right].
\label{gravrule2}
\end{equation}
The first term is the eikonal Feynman rule of eq.~(\ref{eikrules}),
and the remainder is then the NE Feynman rule of
eq.~(\ref{gravrule3}).\\

Now consider that the external line in figure~\ref{fig:ampleg} is a
fermion. Again neglecting an overall factor of $\kappa/2$, combining
the propagator and vertex gives
\begin{equation}
{\cal M}_n\frac{(\slsh{p}_i+\slsh{k})}{4}\frac{[\gamma^{(\mu}(2p_i+k)^{\nu)}-2\eta^{\mu\nu}\gamma^\alpha(2p_i+k)_\alpha]}{(p_i+k)^2}u(p_i),
\label{ferm1}
\end{equation}
where we have explicitly included the spinor associated with the line
$i$. The reason for doing this is that after expanding
eq.~(\ref{ferm1}) to first subleading order in the soft gluon momentum
$k$, we may simplify the result by anticommuting factors of $\slsh{p}$
and using the Dirac equation
\begin{equation}
\slsh{p}_iu(p_i)=0.
\label{Dirac}
\end{equation}
The result is
\begin{equation}
{\cal M}_n\left[\frac{p_i^\mu\,p_i^\nu}{p_i\cdot k}-\frac{p_i^\mu\,p_i^\nu\,k^2}{2(p_i\cdot k)^2}+\frac{p_i^{(\mu}k^{\nu)}}{4p_i\cdot k}+\frac{\slsh{k}\gamma^{(\mu}p_i^{\nu)}}{4p_i\cdot k}-\frac{\eta^{\mu\nu}}{2}\right]u(p_i).
\label{ferm2}
\end{equation}
The first term is the (spin-independent) eikonal Feynman rule of
equation~(\ref{eikrules}). For the remaining terms, we may rewrite the
combination
\begin{equation}
\frac{\slsh{k}\gamma^{(\mu}p_i^{\nu)}}{4p_i\cdot k}=\frac{k^{(\mu}p_i^{\nu)}}{4p_i\cdot k}-\frac{i k_\rho\Sigma^{\rho(\mu}p_i^{\nu)}}{8p_i\cdot k},
\label{ferm3}
\end{equation}
where we have introduced the spin tensor of eq.~(\ref{sigmadef}). One
then finds the NE 1-graviton vertex of eq.~(\ref{gravrule3}).\\

Note that in this appendix we have not contracted the amplitude with
the polarization tensor for the external graviton. This means that the
one-graviton vertices we have obtained are also valid for off-shell
gravitons.

\bibliography{refs.bib}
\end{document}